\documentclass{ceurart}

\usepackage{multirow}
\usepackage{markdown}
\usepackage{subcaption}
\usepackage{multicol}
\usepackage{algorithm}
\usepackage{algpseudocode}

\def\xmm1{\mathbf{x}^{1-m}}

\def\q{\mathbf{q}}

\def\r{\mathbf{r}}

\def\m{\mathbf{m}}
\def\x{\mathbf{x}}
\def\user{\x}
\def\y{\mathbf{y}}
\def\yt{\mathbf{y}_t}

\def\yc{\mathbf{y}_c}

\def\v1{\mathbf{1}}

\def\f{\mathbf{f}}
\def\e{\mathbf{e}}
\def\ec{\mathbf{e}^{c}}
\def\el{\mathbf{e}^{l}}
\def\es{\mathbf{e}^{s}}
\def\ml{\mathbf{m}^l}
\def\mc{\mathbf{m}^c}
\def\ms{\mathbf{m}^s}

\def\LCLXR{\mathcal{L}^\text{CLXR}}
\def\Lpred{\mathcal{L}^\text{pred}}
\def\Lcmp{\mathcal{L}^\text{cmp}}
\def\Lreg{\mathcal{L}^\text{reg}}
\def\lampred{\lambda_\text{pred}}
\def\laminv{\lambda_\text{inv}}

\def\lamcmp{\lambda_\text{cmp}}
\def\lamreg{\lambda_\text{reg}}
\def\Linv{\mathcal{L}^\text{inv}}

\def\items{\mathcal{V}}
\def\users{\mathcal{U}}
\def\rank{\r_{\f}}

\def\mprr{\text{MPRR}}
\def\mprrp{\mprr_{\%}}
\def\mprrh{\mprr_{\#}}

\def\indic{\mathbb{I}}

\begin{document}

\copyrightyear{2026}
\copyrightclause{Copyright for this paper by its authors.
  Use permitted under Creative Commons License Attribution 4.0
  International (CC BY 4.0).}

\conference{IntRS'26: Joint Workshop on Interfaces and Human Decision Making for Recommender Systems, September 28, 2026, Minneapolis.}
\title{Why This, Not That? Mining User Profiles for Pair-wise Counterfactuals}


 \author[1]{Meysam Varasteh}[%
 orcid=0009-0003-0346-4951,
 email=meysam.varasteh@colorado.edu,
 ]
 \cormark[1]
 \address[1]{Department of Computer Science, University of Colorado Boulder,
   Boulder, CO 80309, USA}

 \author[2]{Veronika Bogina}[%
 orcid=0000-0002-8005-7618,
 email=sveron@gmail.com,
 ]
 \address[2]{Tel Aviv University, Tel Aviv, Israel}

 \author[2]{Noam Koenigstein}[%
 orcid=0000-0001-8219-4512,
 email=noamk@tauex.tau.ac.il,
 ]

 \author[3]{Robin Burke}[%
 orcid=0000-0001-5766-6434,
 email=robin.burke@colorado.edu,
 ]
 \address[3]{Department of Information Science, University of Colorado Boulder,
   Boulder, CO 80309, USA}

 \cortext[1]{Corresponding author.}

\begin{abstract}
The topic of explanation in recommender systems has seen steady research attention since the earliest days of the field. With some exceptions, this work has focused on the explanation of single items in a recommendation list and, especially recently, has emphasized approaches that are decoupled from the logic of the recommendation algorithm itself. Based on findings in the psychology of interpersonal communication, we propose a new task, pairwise interpretation of item rankings, asking the comparative question ``Why is item A ranked higher than item B?''. An effective solution to this task, we argue, is inherently grounded in the operation of the recommendation algorithm. We propose a class of techniques based on counterfactual learning to uncover the items in a user's profile that have contributed to the relative ranking of items. Using multiple datasets, we show that it is possible to identify such items as potential basis for comparative explanation.
\end{abstract}


\maketitle

\section{Introduction}
Why was this item recommended? This question has been central to recommender systems research for decades, with well-established methods for explaining the presence of a single item in a ranked list \cite{tintarev2015explaining,tintarev2022beyond}. In this work, we focus on a different and equally natural question: ``Why is item $T$ ranked higher than item $C$?''. This problem is referred to as ``Comparing Item Rankings'' in \cite{varasteh2024comparative}.

Psychological studies show that people often seek contrastive explanations when reasoning about decisions \cite{miller2019explanation}. Comparison is a fundamental component of human judgment \cite{tversky1988contingent}, and preferences can be influenced simply by altering the set of alternatives \cite{simonson1992choice}. In the context of recommender systems, comparative explanations have the potential to enhance clarity and relatability by focusing on the specific factors that distinguish two competing items, rather than enumerating all reasons for recommending $T$ in isolation. This narrower focus can make the explanation both more intuitive for the user and potentially easier to generate algorithmically.

From an algorithmic perspective, the question ``Why is $T$ ranked above $C$?'' is inherently counterfactual: the user implicitly asks ``and not the other way around?''. Addressing this requires identifying the minimal changes to the underlying data that would reverse the relative positions of the two items. Such counterfactual comparative explanations directly expose causal relationships between user data and ranking outcomes, aligning naturally with transparency goals in recommender systems \cite{tintarev2022beyond}. They are valuable not only for end users but also for system designers and auditors, and they support regulatory demands for explanations of how personal data influences automated decisions, such as those in the GDPR\footnote{\url{https://www.consilium.europa.eu/en/policies/data-protection/data-protectionregulation/}}.

As a terminological matter, we note that the terms ``comparative'' and ``contrastive'' are both used somewhat interchangeably in different publications in this area \cite{yang2022comparative, jacovi2021contrastive, malandri2024model, castelnovo2024evaluative}. The term ``contrastive explanation'' has a well-established meaning in the explainable AI literature meaning a counterfactual explanation (CE) for a classification: ``Why was this instance classified in category A instead of category B?'' Rather than trying to expand this established definition to cover our case, we use the term ``comparative'' instead \cite{wang2022not, luss2025cell, waa2018contrastive}. We will use the terms \textit{target} item to refer to the higher ranked item $T$ and \textit{comparative} item to refer to the lower ranked item $C$. 

Following the work of \citet{barkan2024counterfactual}, we adopt a learning-oriented approach to this CE problem. Similar to their method, we develop an \textit{explainer model} that learns how modifications in a user's profile influence the recommendation outcomes. Our findings indicate that this method allows us to identify items in a user's profile that, when removed, would reverse the ranking positions of the target item and the comparative item. Such a set of items we term a \textit{perturbation}. Thus, our intended explanation is of the following form: ``The reason item $T$ is ranked higher than item $C$ is because the user profile included items $x_1$, $x_2$, and $x_3$. If these items were not present, $C$ would have been recommended higher than $T$.''

We highlight two aspects relevant to the quality of our explanation output, drawing from Grice's maxims of communication \cite{grice1991studies}. One is \textit{veracity}, the ``maxim of quality''. It should be the case that $C$ will be ranked above $T$ if the user's profile is edited by removing the identified items and the recommendations are re-generated. The second quality is that the explanation should be succinct, the ``maxim of clarity''. Ideally, the suggested perturbation should be the smallest set of items for which the reversal happens. 

This study represents an initial foray into the development of comparative ranking explanations using this approach. We focus here on the feasibility of generating accurate and succinct counterfactual perturbations. We address the following research questions:

\begin{itemize}
    \item \textbf{RQ1}: Can counterfactual learning be used to efficiently, in terms of perturbation size, approximate profile perturbations for generating contrastive ranking explanations?
    \item \textbf{RQ2}: How does a contrastive objective perform compared to one that decouples target and comparative items?
    \item \textbf{RQ3}: How do the results vary across different datasets and recommendation algorithms?
\end{itemize}

\section{Related Work}
\label{sec:related}
As noted above, explanation is a core topic in recommender systems research, and there are a variety of well-known techniques, especially for explaining the presence of a single item in a recommendation list. Readers are referred to \cite{tintarev2015explaining,tintarev2022beyond} for details of the state of the art in these areas. Our research directly addresses a need identified in \cite{varasteh2024comparative}, which surveys the comparative explanation task and proposes several variations. However, this position paper focuses solely on defining the comparative explanation problem and does not present any implementations.

In the realm of XAI, model-agnostic methods like SHAP~\cite{shap_original} and LIME ~\cite{lime_original} stand out for their versatility and flexibility. These techniques are not tied to any specific model architecture, allowing them to be applied across various ML models, including those used in recommendation systems. SHAP and LIME generate explanations by perturbing input vectors to assess the contribution of each feature to the model's predictions. However, when applied to recommender systems, this approach often requires perturbing the entire user dataset, which can be computationally intensive and impractical for real-time applications with many items.

Counterfactual explanations for recommendations are a more recent innovation, drawing on research from XAI. CEs generally seek to demonstrate how altering specific inputs would affect a model's prediction \citep{verma2024counterfactual}. This approach is useful for revealing the causal links between user data and the recommendations provided. By pinpointing modifications in user data that would lead to different recommendations. Recent examples of these methods include \cite{kaffes2021model,tan2021counterfactual,zhong2022shap}. These techniques are challenging to employ in practical contexts because they require search over a large perturbation space for each user, a process that would be prohibitive in a real-world setting. 

A complementary line of work moves beyond single-instance perturbations by learning a model to predict counterfactual explanations at scale. Rather than searching for perturbations individually, these approaches train a mask over user profiles or item features to generate explanations efficiently across many instances \cite{barkan2024counterfactual, baklanov2025refining, mohammadi2025beyond, verma2024counterfactual, barkan2026fidelity}. In this work, we adapt the LXR framework proposed by \citet{barkan2024counterfactual}, which learns to predict profile masks, for the comparative explanation task.

Several recent papers have addressed the comparative explanation task for ranking  \cite{castelnovo2024evaluative,salimiparsa2023counterfactual,singh2018posthoc} and NLP  task \cite{jacovi2021contrastive,luss2025cell}. In this line of research, the items are compared based on their attributes. For example, in \cite{castelnovo2024evaluative}, the authors provide an example explanation: \textit{Characteristics in favor of Candidate 00079 include a higher score in HSC\_P and a higher score in SSC\_P. Characteristics in favor of Candidate 00188 include a higher score in DEGREE\_P and having previous work experience.} However, this approach is not easily transferable to recommendation tasks, where ranking criteria are personalized. 
In contrast, our counterfactual explanation is more closely tied to the user's profile, focusing on what would need to change in the user's preferences for the recommendation to differ. Also, \cite{jacovi2021contrastive} proposed a method that produces contrastive explanations by projecting the latent space of inputs. However, since one of the main characteristics of our model is being model-agnostic, this approach is not used. 

Contrastive explanation is also applied in Information Retrieval (IR) due to the ranking nature of the models, which are designed to order documents for each query. Several studies have investigated contrastive explanations with the aim of addressing ranking related questions, such as: "Why is this document ranked higher than the other?" For example, \cite{singh2019exs} proposed a variant of the LIME \cite{lime_original}method adapted specifically to the contrastive setting, while \cite{singh2020model} approximates the original model with a simple ranker to explain "Why document $d_i$ is more relevant than $d_j$. Similarly, \cite{rennings2023listwise} incorporated multiple simple rankers to provide listwise explanations for ranking models. However, these IR methods are not applied to cases where ranking is personalized as in recommendation.

There is also research that uses the term ``comparative explanation'' (for example, \cite{yang2022comparative}) to describe explanations that compare a single recommended item against an item or items from the user's profile. This is also a different explanation task than ours, which poses a counterfactual question about two recommended items and requires very different explanation methods.

\section{Explanation Models}
\label{sec:model}
Our approach to comparative explanation builds directly on the LXR framework~\cite{barkan2024counterfactual,gurevitch2025lxr}, a state-of-the-art method for generating counterfactual explanations of individual recommended items. We first outline the LXR method to establish the foundation of our work, then describe how we extend it to address the comparative setting. All notation used throughout the paper is summarized in Table~\ref{tab:notation}.


\begin{table}[bt]
\begin{tabular}{|c|l|}
\hline
\textbf{Symbol} & \textbf{Description} \\ 
\hline
$\users$ & The set of users. \\
\hline
$\items$ & The set of items.\\ 
\hline
$\user$ & The user profile, \\
    & represented as a binary vector over items. \\
\hline
$\yt, \yc$ & The target and comparative items,\\ 
           & represented as a one-hot vectors. \\
\hline
$\m$ & An explanation map / importance scores \\ 
    & (superscript denotes type). \\
\hline
$\e$ & An explainer (superscript denotes type). \\
\hline
$\f$ & A recommender system. \\ 
\hline
$\rank(\user, \y)$ & The rank of item $\y$ in the recommendations \\ 
& produced by $\f$ for user $\user$. \\
\hline
$\Pi$ & A perturbation of a user profile. \\
\hline
\end{tabular}
\caption{Table of notations}
\label{tab:notation}
\end{table}

\subsection{LXR}
LXR~\cite{barkan2024counterfactual} formulates the task of explaining a single recommended item as identifying the specific items in a user's profile $\x$ whose removal would cause that item to disappear from the top slate of recommendations. The approach is learning-based: the explainer is a trainable function optimized via a counterfactual loss over the recommender's outputs.

Formally, the learnable explainer $\el : \{0,1\}^{|\items|} \times \{0,1\}^{|\items|} \rightarrow [0,1]^{|\items|}$ takes as input the user’s interaction history $\x \in \{0,1\}^{|\items|}$ and a target item $\y \in \{0,1\}^{|\items|}$ represented as a one-hot vector, and produces an \emph{explanation mask} $\ml \in [0,1]^{|\items|}$, where each entry represents the estimated contribution of the corresponding profile item to the recommendation of $\y$.

We can order the profile items by their score in the mask and establish some threshold for removal. The set of items removed we call the profile \textit{perturbation}, denoted by $\Pi$. The quality of $\Pi$ is determined by its \textit{veracity}: if the items in $\Pi$ are removed from $\x$, the recommender should cease to recommend $\y$. The conciseness of the explanation is measured by the size of $\Pi$: ideally it includes only a small subset of items whose removal leads to a sharp drop in $\y$’s rank. 



\subsection{Comparative Explanation}
Next, we formulate a \emph{comparative explainer} that tries to explain the recommender's preference for a target item over a comparative item.  Let $\ec:\{ 0,1 \}^{|\items|} \times \{ 0,1 \}^{|\items|} \times \{ 0,1 \}^{|\items|} \rightarrow [0,1]^{|\items|}$ be a comparative explainer that receives a user data vector $\x \in \{ 0,1 \}^{|\items|}$ and two one-hot vectors: one representing the target item $\yt \in \{0,1\}^{\items}$, and another, the comparative item $\yc \in \{0,1\}^{\items}$. Similar to LXR, the output of the explainer is an \emph{explanation mask} $\mc = \e(\x, \yt, \yc)$, where $\mc[i]$ indicates the contribution of item $i$ from the user's history to the recommender's preference of the target item $\yt$ over the comparative item $\yc$. In what follows, we present several approaches to implement such a comparative explainer. 

\subsubsection{CLXR-score:} 
A naive implementation of a comparative explainer can be based on a simple adaptation of a pre-trained LXR explainer $\el$ to the aforementioned comparative task. This can be achieved by considering the cases of the target item and the comparative item independently, using the explanation scores from the LXR explainer. We refer to this variant as \textit{CLXR-score}.

Given a pre-trained LXR explainer $\el$, we employ it to produce two explanation maps $\ml_t$ and $\ml_c$ and combine them as follows:
\begin{align}
    \label{eq:sCLXR}
    \ml_t &= \el(\x, \yt) \\ \nonumber
    \ml_c &= \el(\x, \yc) \\ \nonumber
    \mc &= \ml_t \odot (\mathbf{1} - \ml_c)   
\end{align}
where $\mc$ is the comparative explanation mask, $\mathbf{1}$ is a vector of ones, and $\odot$ is the element-wise product. 

The $\ml_t$ mask identifies the items in the profile that are likely to reduce the rank of $\yt$ if removed from the user profile. The inverse of the $\ml_c$ mask identifies items that do not contribute to the $\yc$'s rank. Therefore, the multiplicative combination of these values focuses $\mc$ on items that will reduce the rank of $\yt$ but not reduce the rank of $\yc$ if removed from the profile. 

\subsection{Learning Comparative Explanations}
The primary advantage of the CLXR-score implementation is its reliance on the pre-trained LXR model from~\citet{barkan2024counterfactual}. This allows a single LXR explainer to be used for both single-item explanations and comparative explanations without additional optimization. Next, we explore solutions that directly optimize for comparative explanations.

Given the user’s history vector $\x$, target item $\yt$, and comparative item $\yc$, our objective consists of four loss terms:  
\begin{itemize}    
    \item $\Lcmp \big( \x, \yt, \yc \big)$: Encourages finding explanations that emphasize the target item while suppressing the comparative item.
    \item $\Lpred \big( \x, \yt \big)$: Inspired by the LXR objective, this term prioritizes supporting the target item.
    \item $\Linv \big( \x, \yt \big)$: This term is designed to suppress the recommendation of the target item when the most important elements related to the target item are removed.

    \item $\Lreg \big( \mc \big)$: Regularization term promoting mask sparsity.
\end{itemize}

The complete CLXR objective is:
\begin{align}
 \label{eq:CLXR_loss}
\LCLXR \big( \x, \yt, \yc \big) = 
 & \lamcmp \Lcmp \big( \x, \yt, \yc \big) + \lampred \Lpred \big( \x, \yt \big) + \\ \nonumber
& \laminv \Linv \big( \x, \yt \big) + \lamreg  \Lreg(\mc) 
\end{align}

This framework allows flexibility in defining the loss functions based on the explainer’s architecture. Below, we present two implementation approaches.

\subsubsection{CLXR-joint Implementation} 
Our first implementation, \textit{CLXR-joint}, employs a Siamese network to jointly optimize all loss terms in
Eq.~\ref{eq:CLXR_loss}. This approach builds on the CLXR-score concept but replaces the LXR explainer $\el$ in Eq.~\ref{eq:sCLXR} with an intermediate Siamese network $\es$, defined as \(\es: \{ 0,1 \}^{|\items|} \times \{ 0,1 \}^{|\items|} \rightarrow [0,1]^{|\items|}\). The comparative explanation mask $\mc$ is then computed as:
\begin{align}
\ms_t &= \es(x,\yt),  \\ \nonumber
\ms_c &= \es(x,\yc), \\ \nonumber
 \mc &= \ms_t \odot ( \mathbf{1}-\ms_c)  \\ \nonumber
 \label{eq:CLXR_mask}
\end{align}

Here, the Siamese network $\es$ generates intermediate masks $\ms_t$ and $\ms_c$, akin to $\ml_t$ and $\ml_c$ from Eq.~\ref{eq:sCLXR}. 
Thus, \textit{CLXR-joint} learns a comparative explanation mask $\mc$ using:
\begin{equation}
    \mc = \ec(\x,\yt,\yc) = \es(\x,\yt) \odot \big( \mathbf{1} - \es(\x,\yc)\big),
    \label{eq:mask}
\end{equation}
where $\es$ is optimized using Eq.~\ref{eq:CLXR_loss}. In essence, this architecture is similar to that of CLXR-score, except that in CLXR-joint, $\es$ is optimized directly for a comparison explanation task, whereas in CLXR-score, $\ec$ is the original LXR explainer, trained on explaining single items. 

Specifically, training $\ec$ is performed via optimizing the following loss terms:
\begin{align}
    \label{eq:CLXR_loss_terms}
    \Lcmp &= - \log \big( \yt \f(\x \odot \mc )- \yc \f(\x \odot \mc ) \big), \\ \nonumber 
    \Lpred &=-\log(\yt \f(\x \odot \ms_t)), \\ \nonumber
    \Linv &=  \log(\yt \f(\x \odot (1-\ms_t))) + \log(\yt \f(\x \odot (1-\mc))),   \\ \nonumber
    \Lreg &= | \mc |.
\end{align} 
The comparative loss $\Lcmp$ maximizes the score gap between $\yt$ and $\yc$ after applying the explanation mask $\mc$ to $\x$. The prediction loss $\Lpred$ ensures $\yt$ remains the top-ranked item by maximizing its score when masked by $\ms_t$. The inverse loss $\Linv$ is designed to suppress the recommendation of the target item when the most important items related to it are removed from the user's profile by inverting $\ms_t$ and $\mc$.  
Finally, the regularization loss $\Lreg$ enforces sparsity in $\mc$. This formulation enables a more targeted optimization for comparative explanations, improving fidelity and interpretability.

We note that employing $\Lcmp$, $\Lpred$ and $\Linv$ requires an ``untraditional'' setup where the recommender model $\f$ is embedded within the learning objective and plays an active role in the optimization process. However, the recommender's weights remain frozen to preserve the integrity of the explanation target. 

\subsubsection{TDLR Implementation}
Our second implementation adopts the Targeted Deviation - Logit Ranking (TDLR) loss, commonly used in adversarial attacks~\cite{croce2020reliable}. Hence we dub this implementation \textit{CLXR-tdlr}. Different from the \textit{CLXR-joint} implementation, CLXR-tdlr does not rely on an intermediate network $\es$. Instead, given both items  ($\yt$ and $\yc$), the comparative explainer $\ec$ directly learns the comparative mask $\mc$.  Additionally, we set $\lambda_{\text{pred}}=0$ and $\lambda_{\text{inv}}=0$, effectively removing $\Lpred$ and $\Linv$ from the objective in Eq.~\ref{eq:CLXR_loss}, relying solely on the comparative loss $\Lcmp$ and regularization term $\Lreg$ defined as  :
\begin{align}
    \label{eq:CLXR_TDLR}
    \Lcmp &= - \frac{\yt \f(\x \odot \mc) - \yc \f(\x \odot \mc)}{\f(\x)[1] - \f(\x)[K] + \epsilon}, \\ \nonumber
    \Lreg &= | \mc |,
\end{align}
where $\f(\x)[1] - \f(\x)[K]$ represents the score difference between the top-ranked item and the item at position $K$. We set $K=10$ and $\epsilon=1 \times 10^{-8}$ as hyperparameters.

Similar to the \textit{CLXR-joint} implementation, the comparative loss $\Lcmp$ in the \textit{CLXR-tdlr} implementation utilizes the recommender model $\f$ as part of the learning objective (without changing $\f$). 
Minimizing $\Lcmp$ guides the explainer $\ec$ to find a mask $\mc$ that increases the score gap between $\yt$ and $\yc$. The denominator normalizes this effect, ensuring loss stability across varying score distributions. The regularization term $\Lreg$ remains an $\ell_1$ sparsity constraint to encourage minimal yet effective explanations.

\subsection{Optimization}
\label{sec:optimization}
To train the CLXR explainers, we generate recommendation slates for each training set user using the recommendation algorithm. We assign each user’s target item $\yt$ as the top-ranked recommendation from this slate. The comparative item is randomly sampled from the remaining items ranked $2$ through $10$.

In both implementations, the explainer networks are modeled as simple feed-forward multi-layer perceptrons. The explainer’s parameters, denoted by $\theta$, are optimized using stochastic gradient descent to minimize the following objective:
\begin{equation}
\theta^* = \arg\min_{\theta} \frac{1}{|\users|} \sum_{u=1}^{|\users|} \LCLXR(\x, \yt, \yc),
\end{equation}
where $\users$ is the set of users, and $\LCLXR$ as in Eq.~\ref{eq:CLXR_loss}.

\section{Evaluation Metrics}
\label{sec:metrics}
Comparative explanation of recommendation outputs is a novel problem, and no established metrics exist for evaluating the quality of potential solutions. We therefore define two metrics to assess the effectiveness of our algorithms. Since evaluating all possible item pairs in a user's recommendation list is impractical, we focus on the top-ranked item for each user, denoted as $\yt$ where $\rank(\x, \yt) = 1$. The comparative item $\yc$ is sampled from the other top-$K$ recommendations, with $K=10$.

The user profile $\x$ is a set of items $x_i$ previously rated by the user. Our algorithms produce explanations in the form of an attribution mask $\mc$ over these items, indicating their estimated influence in ranking $\yt$ above $\yc$. As discussed earlier, our intended counterfactual explanation takes the form: “Item $T$ is ranked higher than item $C$ because you liked $x_1$, $x_2$, and $x_3$. If you had not liked these items, $C$ would have been ranked higher than $T$.”

A perturbation $\Pi_{\sigma}$ is defined as the subset of items removed from the profile to induce rank reversal, where $\sigma$ is the fraction of profile items with the highest attribution scores. For example, $\Pi_{0.2}$ denotes the top 20\% of items in $\x$ according to $\mc$.

We measure the size of the smallest perturbation that reverses the ranks of $\yt$ and $\yc$, with smaller values indicating a better explainer. This is the \emph{Minimal Perturbation for Ranking Reversal} (MPRR):

\begin{equation}
    \mprrp(\x, \yt, \yc) = \min_{\sigma \in [0,1]} \{ \sigma : \f(\x \setminus \Pi_{\sigma}, \yt) < \f(\x \setminus \Pi_{\sigma}, \yc) \}.
    \label{eq:mprr}
\end{equation}

Note that our search over $\sigma$ considers progressively larger subsets of top-ranked items according to $\mc$, rather than all possible item subsets. As a result, it is possible for a valid perturbation to exist but remain undiscovered if its items are not ranked highly by the explainer. Hence, if no perturbation in this sequence produces a rank reversal, we set $\mprrp = \infty$ and exclude that user from the MPRR calculation.

Differences in profile length make percentages more comparable across datasets, but we also report $\mprrh$, the corresponding perturbation size in absolute item counts.

Some rank reversals may be fundamentally impossible, for example due to strong popularity effects in the data. For example, consider two diverse movies: Abbas Kiarostami’s \textit{A Taste of Cherry} and Greta Gerwig’s \textit{Barbie}. While \textit{A Taste of Cherry} is an acclaimed film, even winning the Palme d’Or in 1997, a broad-based recommender system trained on the preferences of
the general American audience is unlikely to rank it above \textit{Barbie}, a blockbuster with extensive marketing and widespread appeal. If a user receives \emph{Barbie} as their top recommendation, no realistic perturbation may push \emph{A Taste of Cherry} above it, given its niche appeal. In such cases, a system might instead need to fall back to a general, non-personalized explanation.

Because not all pairs can be explained, we also report \emph{Coverage}:

\begin{equation}
    \text{Coverage} = \frac{1}{|\users|} \sum_{\x \in \users} \indic \big( \mprrp(\x, \yt, \yc) < \infty \big),
    \label{eq:coverage}
\end{equation}
where $\indic$ is the indicator function.

High Coverage means explanations can be generated for more users, while low $\mprrh$ means explanations are more succinct. We expect a trade-off: more precise perturbations may be possible for fewer users, while broader coverage may require accepting larger perturbations. The appropriate balance depends on the intended application.

\section{Methodology}
\label{sec:method}
\subsection{Data sets}

Our evaluations are based on the MovieLens 1M (ML-1M), Yahoo Music (Yahoo), and Pinterest datasets.
To simulate implicit ratings for ML-1M, we kept only ratings of 3.5 or higher and included only users and items with at least two ratings. The resulting dataset comprised 575,128 ratings from 6,037 users across 3,381 items. For the Yahoo dataset, we included ratings of 70 or higher and retained users and items with at least two ratings. To keep the datasets similar in size, we sampled a subset of 13,797 users, yielding a dataset with 365,750 ratings across 4,604 items. Finally, for the Pinterest dataset \cite{he2017neural}, we randomly sampled 486,744 ratings from 19,155 users covering 9,362 items.
Each dataset was split into 80\% for training and 20\% for testing. 10\% of the training data was used for hyperparameter tuning. The code for data preparation, model training, and evaluation is available via anonymous GitHub\footnote{https://github.com/that-recsys-lab/CLXR-an}.

\subsection{Recommendation models}
Although the explainer model is designed to be algorithm-agnostic, its performance is a function of the choice of recommendation algorithm $\f$, especially concerning the well-known issue of popularity bias \cite{jannach2015recommenders}. To explore this, we present the results of the experiments using two different recommendation algorithms, following the approach in \cite{barkan2024counterfactual}. The performance of two recommendation models across the three datasets is presented in Table \ref{tab:recommender_performance}.

\textbf{Matrix factorization (MF):} Matrix factorization is a widely used technique in recommender systems that models the interaction between users and items to predict preferences. The goal is to decompose the user-item interaction matrix into two lower-dimensional matrices: user latent factors and item latent factors. Each user and item is represented as a vector in a shared latent space\cite{varasteh2023improved}. The predicted interaction is computed as the dot product of these vectors. We developed a version of matrix factorization (MF) where the model takes a binary encoding of the user’s historical interactions and dynamically computes the user’s latent representation using a simple projection matrix.

\textbf{Variational Autoencoder (VAE):} A Variational Autoencoder (VAE) is a probabilistic generative model that has been effectively applied in recommender systems for learning user and item representations. It consists of two main components: an encoder and a decoder. The encoder maps the user’s interaction history into a latent space, producing a probabilistic latent representation modeled as a Gaussian distribution \cite{rendle2022revisiting, rendle2020neural}. The decoder reconstructs the original interaction data from this latent representation. Our VAE recommender model uses a similar architecture to that in  \cite{liang2018variational} and \cite{barkan2024counterfactual}.  
\begin{table}[ht]
\centering
\begin{tabular}{|l|cc|cc|cc|}
\hline
 & \multicolumn{2}{c|}{\textbf{ML1M}} & \multicolumn{2}{c|}{\textbf{Yahoo}} & \multicolumn{2}{c|}{\textbf{Pinterest}} \\
\cline{2-7}
& \textbf{HR} & \textbf{CC} & \textbf{HR} & \textbf{CC} & \textbf{HR} & \textbf{CC} \\
\hline
MF  & 0.08 & 0.13 & 0.23 & 0.20 & 0.04  & 0.13 \\
VAE & 0.19 & 0.44 & 0.40 & 0.23 & 0.1 & 0.72 \\
\hline
\end{tabular}

\caption{Hit rate (HR) and catalog coverage (CC) at 10 for the recommendation algorithms / dataset combinations.}
\label{tab:recommender_performance}
\end{table}
\subsection{Explanation models}
 In this study, we evaluate the following models for ranking items to generate counterfactual comparative explanations.

\textbf{Popularity (POP):} 
This is a simple baseline that ranks items according to their popularity score (i.e., number of associated actions).

\textbf{SHAP:} SHAP (SHapley Additive exPlanations) \cite{lundberg2017unified} is a popular model-agnostic method for explaining predictions of machine learning models, based on the concept of Shapley values. In the context of recommender systems, the number of perturbations grows exponentially with the number of items in a user's profile, making this approach computationally expensive. In \cite{zhong2022shap}, SHAP was applied to recommender systems using only 12 explainable features. 
In our case, we grouped each user's items into $K=10$ clusters and computed Shapley values for the aggregated item clusters.

\textbf{LIME:} 
LIME (Local Interpretable Model-Agnostic Explanations) \cite{ribeiro2016should} is a local, model-agnostic explanation method that approximates the original model using linear surrogate models. In LIME, the original input is perturbed, weighted based on proximity, and used to fit a surrogate model such as linear regression. In \cite{nobrega2019towards}, LIME-RS was introduced for recommender system applications. 

\textbf{ACCENT:} ACCENT (Action-based Counterfactual Explanations for Neural Recommenders for Tangibility) \cite{tran2021counterfactual} is a model-agnostic counterfactual explanation framework based on influence functions for neural networks. It employs the Fast Influence Analysis (FIA) \cite{cheng2019incorporating} mechanism to approximate the contribution of each user's action, based on its estimated impact on the prediction generated by the neural recommender. ACCENT identifies counterfactuals whose removal from the training set would lead to a different recommendation.

\textbf{LXR:}
Although LXR is not specifically designed for comparative explanation, it identifies items that support the recommendation of a single target item. A perturbation involving these items will lower the target’s rank, and may drop the target item below the comparator item. We use the LXR implementation from \cite{barkan2024counterfactual}, and as the results will show, it is competitive with our other methods under some conditions.

\textbf{CLXR-score:}
As described above, CLXR-score extends LXR to the contrastive task by independently handling the target and comparative items and constructing a mask that integrates their scores, as shown in Eq.~\ref{eq:sCLXR}. 

\textbf{CLXR-joint:}
The CLXR-joint model is trained by minimizing the objective function presented in Eq.~\ref{eq:CLXR_loss}. Similar to \cite{barkan2024counterfactual}, the MLP in the explainer comprises $d=20$ hidden layers with dimension $ d$. The output dimensions of the first and second MLP layers are set to $2d$ and $3d$, respectively. The explainer is trained using the Adam optimizer \cite{kingma2017adam} with a learning rate of 0.001 and a batch size that varies depending on the dataset and recommendation model. 

\textbf{CLXR-tdlr:}
The CLXR-tdlr model is trained to minimize the objective function described in Eq.~\ref{eq:CLXR_TDLR}. It has the same structure as CLXR-joint and was trained and tuned using the same procedures.

We performed a grid search to optimize all hyperparameters as well as batch size, learning rate, and latent factor dimensionality. Optimal hyperparameters can be found in the repository.  

\subsection{Experimental procedures}
The experiments can be divided into two stages: the training of the explainer model and its evaluation. The first task is to assemble the training data for the explainer. After training the recommender system $\f$, we generate recommendations for all users. For each set of recommendations, we select $\yt$ (item of rank 1) and $\yc$ (randomly chosen from ranks 2-10). The explainer training data therefore consists of $<\user, \yt, \yc>$ tuples across the training set. The explainer $\ec$ is trained over this data by optimizing the objective described in Section \ref{sec:optimization}.

The evaluation stage proceeds similarly, using $\f$ to generate a recommendation list for each test user and identifying $\yt$ and $\yc$ items from these recommendations. Using the explainer model $\ec$, we compute the mask $\mc$, scoring each item in the user profile by its predicted contribution to the relative ranking of $\yt$ and $\yc$. 

Given the mask $\mc$, we generate a series of successively larger perturbations $\Pi_{0.1..1.0}$ in $0.1$ increments and use these to generate new recommendations based on perturbed profiles  $\user \backslash  \Pi_\sigma$. We examine the predicted ratings assigned to $\yt$ and $\yc$ in these recommendations in order to compute $\mprrp$ and Coverage.

\section{Results}
\label{sec:results}

\begin{table*}[htbp]
\small
\centering
\begin{tabular}{l|l|cc|cc|cc}
\toprule
\textbf{Recommender} &
\textbf{Method} & \multicolumn{2}{c|}{\textbf{ML-1M}} & \multicolumn{2}{c}{\textbf{Yahoo}} &\multicolumn{2}{c}{\textbf{Pinterest}} \\
             & & Coverage $\uparrow$ & $\mprrh$ $\downarrow$ &
                Coverage $\uparrow$ & $\mprrh$ $\downarrow$ & Coverage $\uparrow$ & $\mprrh$ $\downarrow$ \\
\midrule
MF & Popularity    & 56\% & 50 &  66\% & 15 &62\%& 16\\
 & LIME    & 59\% & 37   & 67\%    &11 &60\%&12 \\
 & ACCENT    & 68\%  & 28 & 79\%  & 11&58\%&14  \\
 & SHAP    &   32\%   & 52  & 63\% &19 &57\%& 16\\
& LXR          & 72\% &   33   &   78\% & 11 &74\%&11 \\
& CLXR-score  & 65\% & 31 &  63\% & 10 &64\%& 11\\
& CLXR-joint (our)   & \textbf{80\%} & \textbf{24} &    \textbf{81\%} &  \textbf{7}&\textbf{77}\%&\textbf{9} \\
& CLXR-tdlr (our)    &   79\%     &    26  &   80\%      &   10  &71\%& 9   \\
\midrule
VAE & Popularity   & 74\% & 64 &   65\% & 20 &72\%&15 \\
 & LIME    & 61\% &  38 & 56\%  &14 &68\%& 14   \\
 & ACCENT    & 77\%   & 34   & 67\% & 12 &\textbf{86}\%&\textbf{9} \\
 & SHAP    & 57\% & 68 &  50\%  & 24 &50\%&22 \\
& LXR          & 82\% & 30 & 73\% & 14 &74\%&15\\
& CLXR-score    & 82\% & 30 &    73\%   &    12   & 77\% & 13   \\
& CLXR-joint (our)   & \textbf{92}\% & \textbf{21} &   \textbf{83}\% & \textbf{9} &78\%&12 \\
& CLXR-tdlr (our)    &   85\%     &    27     &   70\%     & 14   &70\%& 13     \\

\bottomrule

\end{tabular}
\caption{Coverage and MPRR scores for the MF and VAE recommenders on the three datasets. The best result is shown in \textbf{bold}.}
\label{tab:full_results}
\end{table*}
Table \ref{tab:full_results} presents the Coverage and average $\mprr$ scores for the ML-1M, Yahoo and Pinterest datasets using two recommendation models. Higher Coverage values indicate better performance, whereas lower MPRR values are preferable. The results show that CLXR-joint consistently outperforms all baselines in both Coverage and MPRR except for the VAE model on the Pinterest dataset. CLXR-tdlr is also competitive, especially for Matrix Factorization. Additionally, ACCENT, which is based on the Fast Influence Analysis (FIA) on the gap score between the target and comparative items, proves to be the best model for the VAE/Pinterest task and also has good performance in the MF/Yahoo condition. As noted earlier, the model's inherent contrastive structure makes it a suitable baseline.

\begin{figure*}[htbp] 
    \centering
    \begin{subfigure}{0.32\textwidth}
        \centering
        \includegraphics[width=\textwidth]{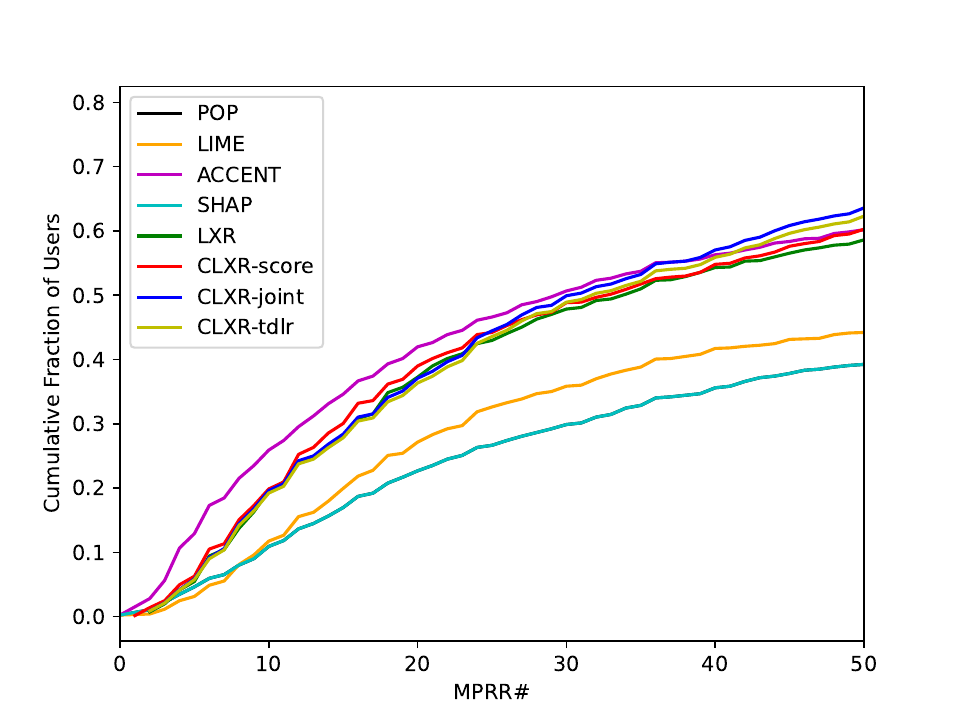} 
        \caption{Recommender: MF, Dataset: MovieLens}
        \label{fig:mf-ml}
    \end{subfigure}
    \hfill
    \begin{subfigure}{0.32\textwidth}
        \centering
        \includegraphics[width=\textwidth]{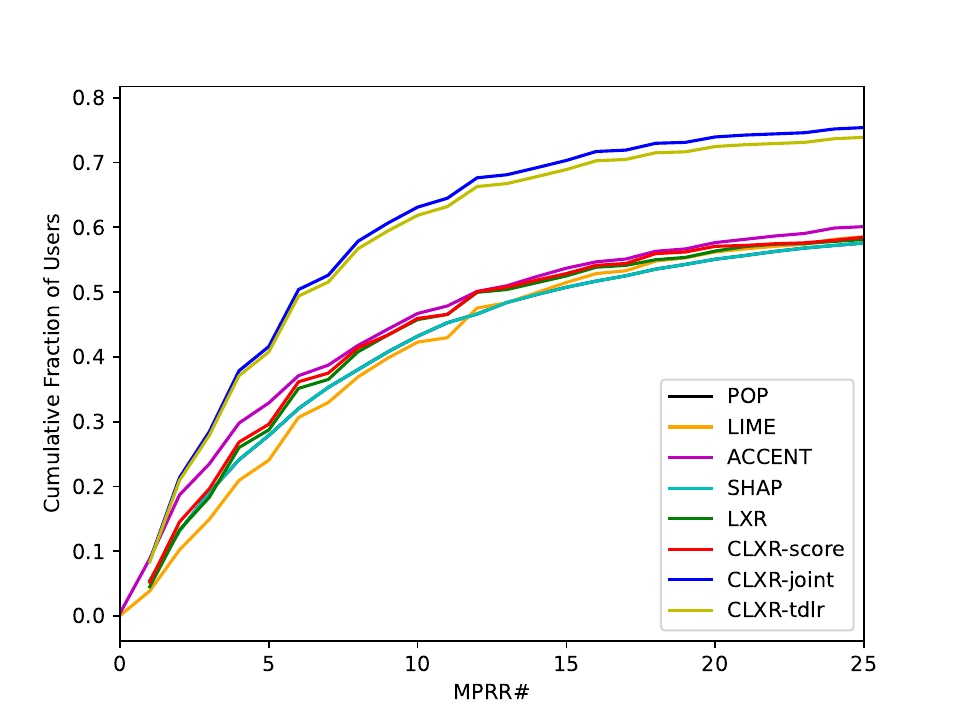} 
        \caption{Recommender: MF, Dataset: Yahoo}
        \label{fig:mf-yh}
    \end{subfigure}
    \hfill
    \begin{subfigure}{0.32\textwidth}
        \centering
        \includegraphics[width=\textwidth]{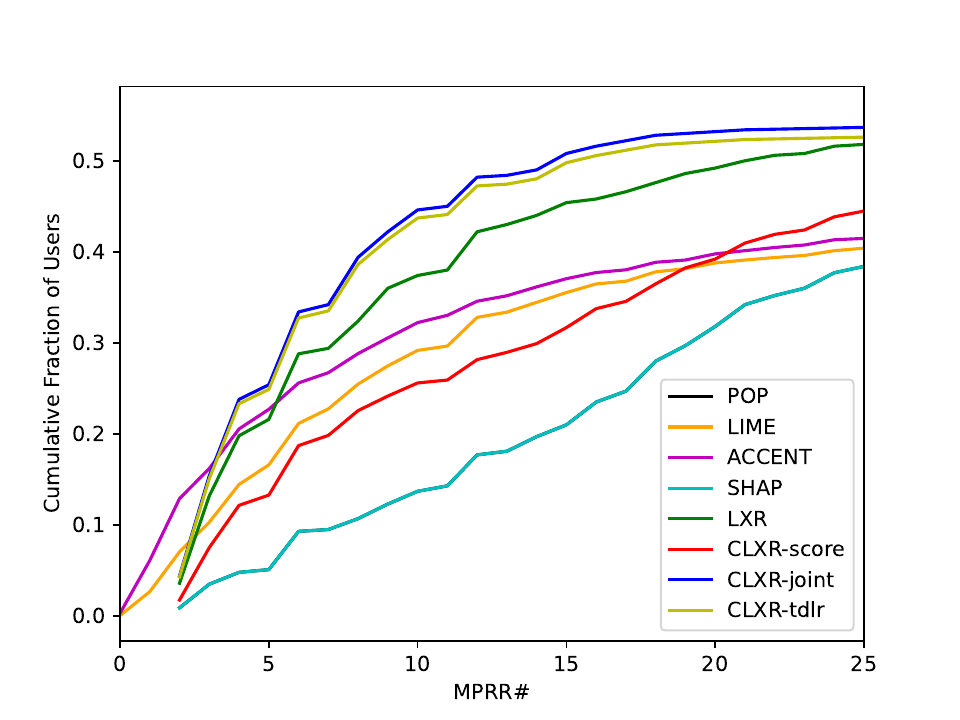} 
        \caption{Recommender: MF, Dataset: Pinterest}
        \label{fig:mf-pin}
    \end{subfigure}

    \vspace{0.2cm}

    \begin{subfigure}{0.32\textwidth}
        \centering
        \includegraphics[width=\textwidth]{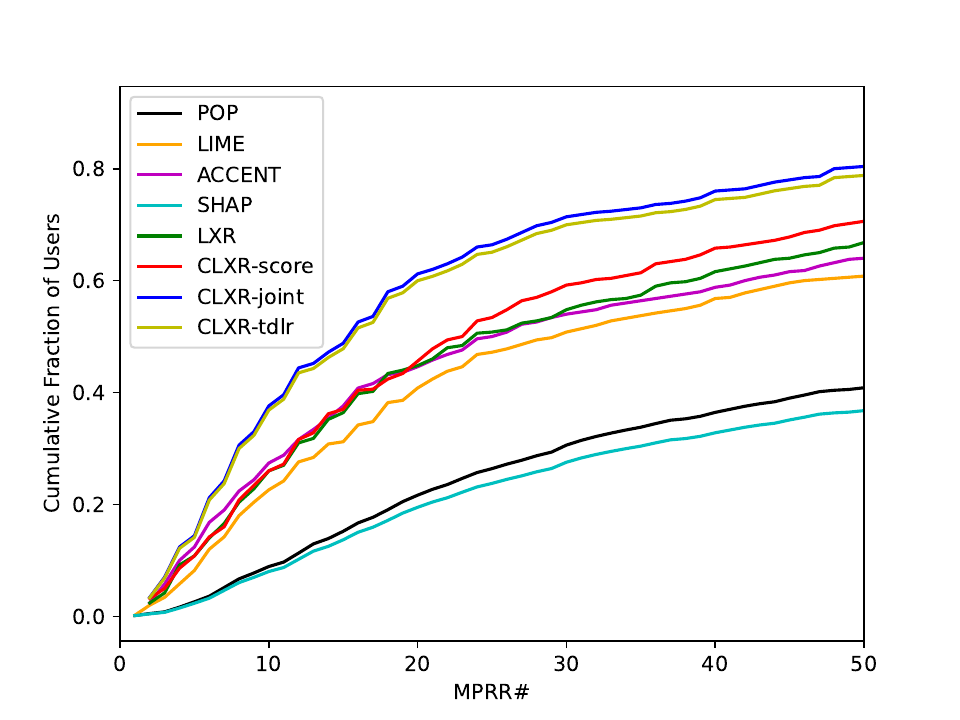} 
        \caption{Recommender: VAE, Dataset: MovieLens}
        \label{fig:vae-ml1m}
    \end{subfigure}
    \hfill
    \begin{subfigure}{0.32\textwidth}
        \centering
        \includegraphics[width=\textwidth]{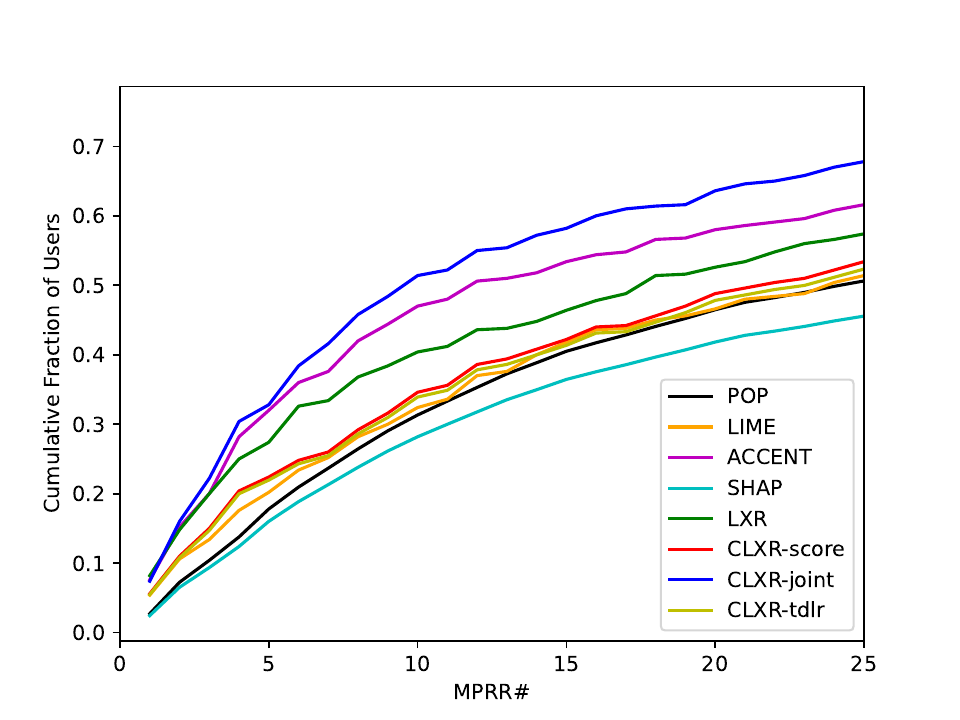} 
        \caption{Recommender: VAE, Dataset: Yahoo}
        \label{fig:vae-yh}
    \end{subfigure}
    \hfill
    \begin{subfigure}{0.32\textwidth}
        \centering
        \includegraphics[width=\textwidth]{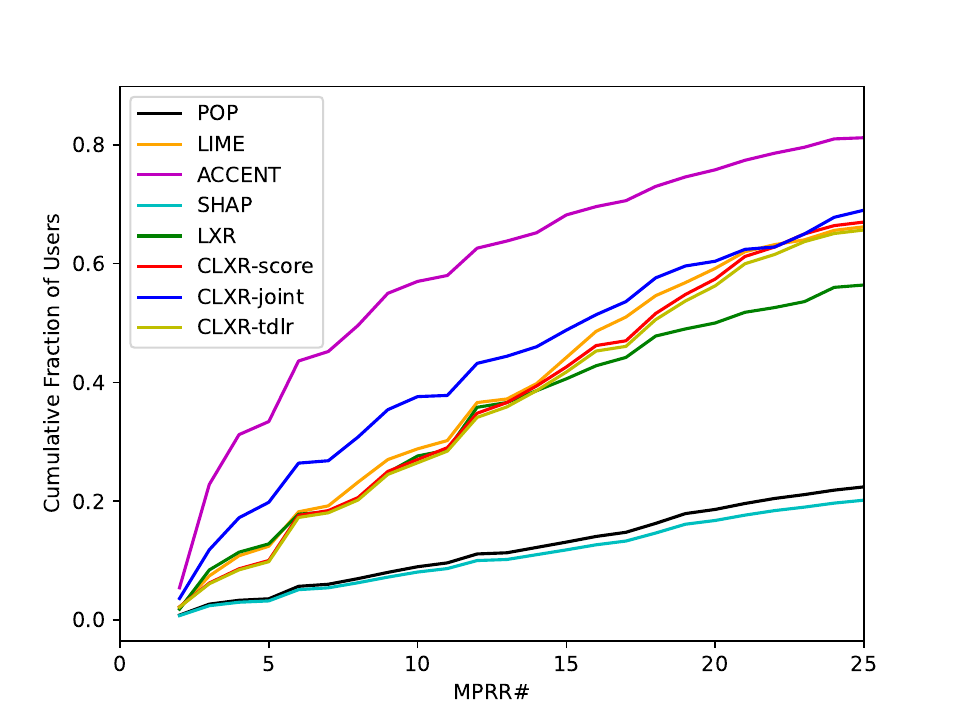} 
        \caption{Recommender: VAE, Dataset: Pinterest}
        \label{fig:vae-pin}
    \end{subfigure}

    \caption{Coverage versus minimum perturbation size for the different algorithms.}
    \label{fig:cumulative_plots}
\end{figure*}

Figure ~\ref{fig:cumulative_plots} illustrates cumulative plots for each explanation model tracking when the target and comparative items are reversed within a perturbation of at most that size. The $Y$-axis represents the cumulative fraction of users, while the $X$-axis indicates the number of masked items or MPRR. We focus on smaller perturbation sizes ($< 50$ for MovieLens, $< 25$ for Yahoo and Pinterest datasets) as these are more likely to be practically useful as explanations.
The most desirable region in these plots is the upper-left quadrant, where an explainer achieves high Coverage (affecting a large number of users) with minimal perturbation to user profiles (MPRR, measuring explanation succinctness). 

These figures provide a more detailed view of algorithm performance while reaffirming CLXR-joint and CLXR-tdlr as the strongest performers except for the VAE/Pinterest. For the matrix factorization model on MovieLens (Figure~\ref{fig:mf-ml}), the curves for all algorithms are relatively close, but CLXR-joint and CLXR-tdlr are dominant throughout. With the VAE recommender on this dataset, CLXR-joint is even more dominant with CLXR-tdlr a bit lower. Note that in these figures, the perturbation sizes are limited to $<50$ and $<25$, whereas in Table~\ref{tab:full_results} there are no such limitations.
When applied to the Yahoo dataset Figure ~\ref{fig:mf-yh}, CLXR-joint and CLXR-tdlr are more clearly dominant. In most conditions, LXR and CLXR-score have similar performance, with LXR actually improving on CLXR-score in Fig~\ref{fig:mf-ml}. 

On the Pinterest dataset with the MF model, CLXR-joint and CLXR-tdlr perform similarly and are the dominant methods. However, for the VAE model, the ACCENT method outperforms all other models, with CLXR-joint as the second best baseline.

\subsection{Explanation Example}
\begin{figure}[htbp]
    \centering
    \includegraphics[width=0.5\textwidth]{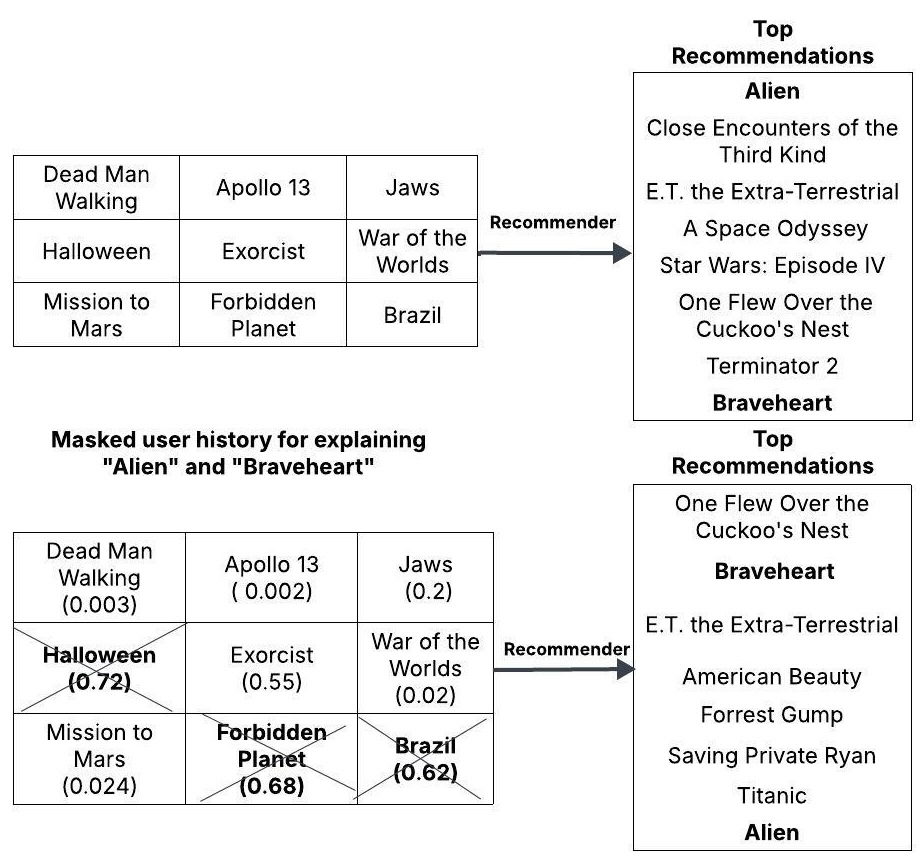} %
  \caption{Example of comparative explanation generation. The original profile and recommendations appear at the top, with \textit{Alien} as the target item and \textit{Braveheart} as the comparative item. The perturbed profile and new recommendations are shown at the bottom.}
  \label{fig:example}
\end{figure}

An example counterfactual perturbation is shown in Figure~\ref{fig:example}. The user profile shown is synthetic, but the perturbation was generated using the CLXR-joint explainer trained on the VAE recommender and the ML1M dataset. The table in the upper left presents the user's profile, while the list on the right displays the corresponding recommendations. The task is to explain why the science fiction/horror classic \textit{Alien} is ranked higher than Mel Gibson's historical epic \textit{Braveheart}. 

The lower figure presents the perturbed user profile, where \textit{Halloween}, \textit{Forbidden Planet}, and \textit{Brazil} have been removed. The numerical values in the lower table are from the $\mc$ mask, representing the importance of each item to this task as output by the $\ec$ explainer. The revised recommendation list (shown in the lower right) results in a rank reversal between \textit{Alien} and \textit{Braveheart}. From this information, it would be possible to generate an explanation as follows: "If you had not liked \textit{Halloween}, \textit{Forbidden Planet}, and \textit{Brazil}, then I would have recommended \textit{Braveheart} higher than \textit{Alien}." Interestingly, other science fiction films such as \textit{Apollo 13}, \textit{Mission to Mars}, and \textit{War of the Worlds} have much lower $\mc$ scores. 

\subsection{The Impact of Popularity}
Recommendation and ranking systems are known to suffer from popularity bias, which arises because errors on popular items will be penalized more heavily in training as they affect more users than rare items~\cite{abdollahpouri2019impact, abdollahpouri2021user}. Given the performance of the POP algorithm in our experiments, we believe that popularity bias may also impact our explainer models as well and lower their effectiveness. We would expect that a user not liking an item which is generally very popular would have a big impact on their placement in the latent space used to generate recommendations. For example, consider a perturbation that calls for the removal of \textit{Star Wars}, \textit{Titanic}, and other extremely popular movies. We would expect this would have a big impact on the ranking of all items, so perhaps it is counterfactually correct. But at the same time, as an explanation, a list of such items might not make any sense to the user. How helpful is an explanation that tells the user that, if they had not watched these extremely popular items, the ranking of their recommendations might be different? They might find this counterfactual extremely unlikely.  

This line of reasoning suggests that popular items might not be the best items for counterfactual explanation. To study this phenomenon, we created a version of CLXR-joint in which the explainer is trained on filtered user profiles in which the top 15\% most popular items are removed. We call this version \textit{CLXR-filter}. The recommendation model is unchanged. 

Table~\ref{tab:popularity_comparison} shows the Coverage and MPRR results for both datasets and both recommenders. The most notable effect is the drop in coverage. This indicates that popular items are forming a large part of many of the perturbations CLXR-joint is producing. It is possible these would be considered low quality explanations in the eyes of users. There are small MPRR increases, except in the VAE / Yahoo condition, meaning that when explanations can be found, the system is finding similar size perturbations among the lower popularity items. These results suggest that item popularity does interact with our counterfactual explanation model but that the model can still function even without popular items that might be less convincing as explanations. Further studies, especially with human subjects, will be needed to understand how popularity interacts with the perceived quality of perturbation-based explanations.

\begin{table}[tbh]
\centering
\renewcommand{\arraystretch}{1.5} 
\setlength{\tabcolsep}{6pt}     
\resizebox{0.5\textwidth}{!}{ 
\begin{tabular}{|c|c|c|c|c|c|}
\hline
\textbf{Recommender} & \textbf{Explainer} & \multicolumn{2}{c|}{\textbf{ML-1M}} & \multicolumn{2}{c|}{\textbf{Yahoo}} \\ \hline
                        &               & Coverage $\uparrow$ & $\mprrp$ $\downarrow$ & Coverage $\uparrow$ & $\mprrp$ $\downarrow$ \\ \hline
\multirow{2}{*}{MF} & CLXR-joint  & 81\% &  25.3\% & 87.3\%  & 25\% \\ \cline{2-6} 
                     & CLXR-filter & 56.3\%  & 25.8\%  & 24\%  & 27\%  \\ \hline
\multirow{2}{*}{VAE} & CLXR-joint  & 93.3\%  & 22.3\%  & 80.5\% & 30\% \\ \cline{2-6} 
                     & CLXR-filter & 81.1\%  & 22.5\%  & 26\%  & 24\%  \\ \hline
\end{tabular}
}
\caption{Comparison of CLXR-joint (as above) and CLXR-filter}
\label{tab:popularity_comparison}
\end{table}


\section{Conclusion and Future Work}
\label{sec:conclusion}

In this paper, we tackle the recommendation explanation problem of ``Comparing Item Rankings'' from \cite{varasteh2024comparative}. Starting from the LXR counterfactual explanation technique originally developed for explaining single items in recommendation list \cite{barkan2024counterfactual}, we developed a family of techniques for the comparative explanation task. Of these, we show that a combined objective with a Siamese network has the best performance. 

To address \textbf{RQ1}, we explored whether counterfactual learning can be leveraged to approximate profile perturbations that yield meaningful contrastive ranking explanations. Building upon the LXR framework originally designed for explaining single-item recommendations \cite{barkan2024counterfactual}, we extend the method to the comparative setting introduced in \cite{varasteh2024comparative}. Our results show that counterfactual profile perturbations are indeed effective in generating contrastive explanations, and that a model trained with a combined contrastive objective in a Siamese network architecture performs particularly well in approximating the necessary changes to reverse item rankings. An important open challenge in this context is the \textit{optimality} of the generated perturbations. While our current method provides efficient approximations, future work should investigate the gap between these and truly minimal explanations, potentially via brute-force analysis on small-scale datasets.

We showed that explicitly optimizing for comparative explanations leads to improved performance, answering \textbf{RQ2}. By jointly optimizing for the preference between target and comparative items, CLXR-joint produces more focused and faithful explanation masks than the methods that are not optimized for a contrastive loss. 

\textbf{RQ3} focuses on the generalizability of our methods across different datasets and recommendation algorithms. Our experiments on multiple benchmark datasets revealed consistent overall trends, but also uncovered some dataset-specific behaviors. For example, popularity effects are more evident in cases where target items are frequently top-ranked, suggesting that fixed target positions may introduce certain biases. These observations highlight the need for further investigation into how the rank of the target item and the characteristics of the underlying algorithm influence explanation behavior.



Finally, while our current study is focused on technical and empirical evaluation, understanding the quality and usability of comparative explanations from a user perspective remains a critical next step. The off-line metrics developed here help us understand the performance of our different explainer models in technical terms, but these need to be complemented with human-centered evaluation. We are planning a follow-up user study to assess whether the explanations produced by these methods are readily comprehensible, actionable, and aligned with the types of "why this, not that" questions real users might ask. 

As we have seen, counterfactual perturbations must sometimes be large in order to achieve the desired reversal of rank --- affecting dozens of items rather than just a few. We expect that techniques such as clustering will help generalize over larger perturbations. For instance, an explanation summarizing over a large set of items might state: ``If you had not watched 25 Marvel action films, we would not have recommended \textit{Aquaman and the Lost Kingdom} over \textit{Batman Returns}.'' Implementing such explanations would necessitate additional summarization mechanisms, which we plan to explore in future work. 





\clearpage

\bibliography{clxr}

\end{document}